# Some Possible Features of General Expressions for Lovelock Tensors and for the Coefficients of Lovelock Lagrangians Up to the $15^{th}$ Order in Curvature (and Beyond)

C. C. Briggs

*Center for Academic Computing, Penn State University, University Park, PA 16802*

**Abstract.** This paper presents some possible features of general expressions for Lovelock tensors and for the coefficients of Lovelock Lagrangians up to the $15^{th}$ order in curvature (and beyond) in terms of the Riemann-Christoffel and Ricci curvature tensors and the Riemann curvature scalar for $n$-dimensional differentiable manifolds having a general linear connection.

> "The sheer size of many of these calculations boggles the mind. At times, several million terms are manipulated in such calculations, whereas the final result is often quite small."
>
> — Boyle, A., and B. F. Caviness, *Future Directions for Research in Symbolic Computation*, Society for Industrial and Applied Mathematics, Philadelphia, PA (1990), p. 42.

This paper presents an assortment of possible features of general expressions for Lovelock tensors $G_{(p)a}{}^{b}$ and the coefficients $L_{(p)}$ of Lovelock Lagrangians up to the $15^{th}$ order in curvature (and beyond) in terms of the Riemann-Christoffel and Ricci curvature tensors $R_{abc}{}^{d}$ and $R_{a}{}^{b}$, respectively, and the Riemann curvature scalar $R$ for $n$-dimensional differentiable manifolds having a general linear connection, the features having been obtained, for $0 \le p \le 5$, by inspection of complete expressions presented elsewhere[1] and, for $p \ge 6$, by extrapolation of the results obtained for $0 \le p \le 5$.

In accordance with various general definitions given by Müller-Hoissen[2] and Verwimp,[3] general expressions for $L_{(p)}$ and $G_{(p)a}{}^{b}$ are given by

$$L_{(p)} = \begin{cases} 1, & \text{if } p = 0 \\ \frac{(2p)!}{2^{p}} R_{[i_1 i_2}{}^{i_1 i_2} R_{i_3 i_4}{}^{i_3 i_4} \ldots R_{i_{2p-1} i_{2p}]}{}^{i_{2p-1} i_{2p}}, & \text{if } p > 0 \end{cases} \tag{1}$$

and

$$G_{(p)a}{}^{b} = \begin{cases} \delta^{b}_{a}, & \text{if } p = 0 \\ \frac{(2p+1)!}{2^{p+1} p} \delta^{b}_{[a} R_{i_1 i_2}{}^{i_1 i_2} R_{i_3 i_4}{}^{i_3 i_4} \ldots R_{i_{2p-1} i_{2p}]}{}^{i_{2p-1} i_{2p}}, & \text{if } p > 0 \end{cases}, \tag{2}$$

respectively, which formulas comprise $(2p)!$ and $(2p+1)!$ covariant index permutations, respectively, where $\delta^{b}_{a}$ is the Kronecker delta; $R_{ab}{}^{cd} = g^{ce} R_{abe}{}^{d}$; and $g^{ab}$ is the contravariant metric tensor (the possibly non-vanishing covariant derivative of which vanishes identically if the connection under consideration is Riemannian), which satisfies the equation $g^{ac} g_{bc} = \delta^{a}_{b}$, where $g_{ab}$ is the covariant metric tensor.

Following Schouten,[4] the Riemann-Christoffel curvature tensor $R_{abc}{}^{d}$ is defined by

$$R_{abc}{}^{d} \equiv 2 \left(\partial_{[a} \Gamma_{b]}{}^{d}{}_{c} + \Gamma_{[a}{}^{d}{}_{|e|} \Gamma_{b]}{}^{e}{}_{c} + \Omega_{a}{}^{e}{}_{b} \Gamma_{e}{}^{d}{}_{c}\right) \tag{3}$$

using anholonomic coordinates (of which holonomic coordinates are a special case), where $\partial_{a}$ is the Pfaffian derivative, $\Gamma_{a}{}^{b}{}_{c}$ the connection coefficient, and $\Omega_{a}{}^{b}{}_{c}$ the object of anholonomity; the Ricci curvature tensor $R_{a}{}^{b}$ by

$$R_{a}{}^{b} \equiv R_{ca}{}^{bc} = -R_{ac}{}^{bc} = -R_{ca}{}^{cb} + 2\left(\nabla_{[c} Q_{a]}{}^{bc} + S_{ca}{}^{d} Q_{d}{}^{bc}\right), \tag{4}$$

where $\nabla_{a}$ is the covariant derivative, $Q_{a}{}^{bc}$ the nonmetricity tensor, and $S_{ca}{}^{d}$ the torsion tensor; and the Riemann curvature scalar $R$ by

$$R \equiv R_{a}{}^{a} = R_{ba}{}^{ab} = -R_{ab}{}^{ab} = R_{ab}{}^{ba} = -R_{ba}{}^{ba}. \tag{5}$$

Now let

$$a_{p} = \text{number of terms in general expression for } L_{(p)}, \tag{6}$$

$$c_{p} = \text{number of partitions of } p, \tag{7}$$

$$a'_{p} = a_{p} - c_{p}, \tag{8}$$

and

$$b_{p} = \text{number of terms in general expression for } G_{(p)a}{}^{b}. \tag{9}$$

Values for $a_{p}$, $c_{p}$, $a'_{p}$, and $b_{p}$ for $0 \le p \le 5$ determined empirically appear below in Tables (1) through (4).

Table (1). $0^{th}$ through $5^{th}$ Differences of $a_{p}$ for $0 \le p \le 5$

| | | | | | | |
|---|---|---|---|---|---|---|
| $p$ | 0 | 1 | 2 | 3 | 4 | 5 |
| $a_p$ | **1** | **1** | **3** | **8** | **25** | **85** |
| $\Delta a_p$ | 0 | 2 | 5 | 17 | 60 | |
| $\Delta^2 a_p$ | 2 | 3 | 12 | 43 | | |
| $\Delta^3 a_p$ | 1 | 9 | 31 | | | |
| $\Delta^4 a_p$ | 8 | 22 | | | | |
| $\Delta^5 a_p$ | 14 | | | | | |

Table (2). $0^{th}$ through $5^{th}$ Differences of $c_{p}$ for $0 \le p \le 5$

| | | | | | | |
|---|---|---|---|---|---|---|
| $p$ | 0 | 1 | 2 | 3 | 4 | 5 |
| $c_p$ | **1** | **1** | **2** | **3** | **5** | **7** |
| $\Delta c_p$ | 0 | 1 | 1 | 2 | 2 | |
| $\Delta^2 c_p$ | 1 | 0 | 1 | 0 | | |
| $\Delta^3 c_p$ | −1 | 1 | −1 | | | |
| $\Delta^4 c_p$ | 2 | −2 | | | | |
| $\Delta^5 c_p$ | −4 | | | | | |

Table (3). $0^{th}$ through $5^{th}$ Differences of $a'_{p} = a_{p} - c_{p}$ for $0 \le p \le 5$

| | | | | | | |
|---|---|---|---|---|---|---|
| $p$ | 0 | 1 | 2 | 3 | 4 | 5 |
| $a'_p$ | **0** | **0** | **1** | **5** | **20** | **78** |
| $\Delta a'_p$ | 0 | 1 | 4 | 15 | 58 | |
| $\Delta^2 a'_p$ | 1 | 3 | 11 | 43 | | |
| $\Delta^3 a'_p$ | 2 | 8 | 32 | | | |
| $\Delta^4 a'_p$ | 6 | 24 | | | | |
| $\Delta^5 a'_p$ | 18 | | | | | |

Table (4). $0^{th}$ through $5^{th}$ Differences of $b_{p}$ for $0 \le p \le 5$

| | | | | | | |
|---|---|---|---|---|---|---|
| $p$ | 0 | 1 | 2 | 3 | 4 | 5 |
| $b_p$ | **1** | **2** | **7** | **26** | **115** | **596** |
| $\Delta b_p$ | 1 | 5 | 19 | 89 | 481 | |
| $\Delta^2 b_p$ | 4 | 14 | 70 | 392 | | |
| $\Delta^3 b_p$ | 10 | 56 | 322 | | | |
| $\Delta^4 b_p$ | 46 | 266 | | | | |
| $\Delta^5 b_p$ | 220 | | | | | |

In Table (3) for $a'_{p}$, it can be seen that

$$\Delta^{3} a'_{p} = 2^{2p+1} \tag{10}$$

where $0 \le p \le 2$, which yields the recurrence

$$a'_{p+3} = 3 a'_{p+2} - 3 a'_{p+1} + a'_{p} + 2^{2p+1}, \tag{11}$$

where $0 \le p \le 2$. In addition, Eq. (10) yields the formula

$$\Delta^{4} a'_{p} = 3 \Delta^{3} a'_{p}, \tag{12}$$

where $0 \le p \le 1$, which yields the recurrence

$$a'_{p+4} = 7 a'_{p+3} - 15 a'_{p+2} + 13 a'_{p+1} - 4 a'_{p}, \tag{13}$$

where $0 \le p \le 1$.

In Table (4) for $b_{p}$, it can be seen that

$$b_{p+2} = 7 \Delta b_{p} - 9p, \tag{14}$$

where $0 \le p \le 3$, which yields the recurrence

$$b_{p+2} = 7 b_{p+1} - 7 b_{p} - 9p, \tag{15}$$

where $0 \le p \le 3$. In addition, Eq. (14) yields the formula

$$\Delta^{2} b_{p+2} = 7 \Delta^{3} b_{p}, \tag{16}$$

where $0 \le p \le 2$, which yields the recurrence

$$b_{p+4} = 9 b_{p+3} - 22 b_{p+2} + 21 b_{p+1} - 7 b_{p}, \tag{17}$$

where $0 \le p \le 1$.

---

[1] Briggs, C. C., "A General Expression for the Quartic Lovelock Tensor," Preprint gr-qc/9703074; "A General Expression for the Quintic Lovelock Tensor," Preprint gr-qc/9607033.

[2] Müller-Hoissen, F., "Spontaneous Compactification with Quadratic and Cubic Curvature Terms," *Phys. Lett.*, **163B** (1985) 106; "From Chern-Simons to Gauss-Bonnet," *Nucl. Phys. B*, **346** (1990) 235.

[3] Verwimp, T., "On higher dimensional gravity: the Lagrangian, its dimensional reduction and a cosmological model," *Class. Quantum Grav.*, **6** (1989) 1655.

[4] Schouten, J. A., *Ricci-Calculus, Second Edition*, Springer-Verlag, Berlin, Germany (1954), pp. 144-145 and 172.

Values for $a_p$, $c_p$, $a'_p$, and $b_p$ for $0 \le p \le 15$ calculated by means—as needs be—of either of the foregoing recurrences for $a'_p$ and for $b_p$ appear below in Tables (5) through (8).

### TABLE (5). $0^{th}$ THROUGH $15^{th}$ DIFFERENCES OF $a_p$ FOR $0 \le p \le 15$

| $p$ | 0 | 1 | 2 | 3 | 4 | 5 | 6 | 7 | 8 | 9 | 10 | 11 | 12 | 13 | 14 | 15 |
|---|---|---|---|---|---|---|---|---|---|---|---|---|---|---|---|---|
| $a_p$ | **1** | **1** | **3** | **8** | **25** | **85** | **318** | **1234** | **4884** | **19458** | **77727** | **310761** | **1242853** | **4971151** | **19884270** | **79536639** |
| $\Delta a_p$ | 0 | 2 | 5 | 17 | 60 | 233 | 916 | 3650 | 14574 | 58269 | 233034 | 932092 | 3728298 | 14913119 | 59652369 | |
| $\Delta^2 a_p$ | 2 | 3 | 12 | 43 | 173 | 683 | 2734 | 10924 | 43695 | 174765 | 699058 | 2796206 | 11184821 | 44739250 | | |
| $\Delta^3 a_p$ | 1 | 9 | 31 | 130 | 510 | 2051 | 8190 | 32771 | 131070 | 524293 | 2097148 | 8388615 | 33554429 | | | |
| $\Delta^4 a_p$ | 8 | 22 | 99 | 380 | 1541 | 6139 | 24581 | 98299 | 393223 | 1572855 | 6291467 | 25165814 | | | | |
| $\Delta^5 a_p$ | 14 | 77 | 281 | 1161 | 4598 | 18442 | 73718 | 294924 | 1179632 | 4718612 | 18874347 | | | | | |
| $\Delta^6 a_p$ | 63 | 204 | 880 | 3437 | 13844 | 55276 | 221206 | 884708 | 3538980 | 14155735 | | | | | | |
| $\Delta^7 a_p$ | 141 | 676 | 2557 | 10407 | 41432 | 165930 | 663502 | 2654272 | 10616755 | | | | | | | |
| $\Delta^8 a_p$ | 535 | 1881 | 7850 | 31025 | 124498 | 497572 | 1990770 | 7962483 | | | | | | | | |
| $\Delta^9 a_p$ | 1346 | 5969 | 23175 | 93473 | 373074 | 1493198 | 5971713 | | | | | | | | | |
| $\Delta^{10} a_p$ | 4623 | 17206 | 70298 | 279601 | 1120124 | 4478515 | | | | | | | | | | |
| $\Delta^{11} a_p$ | 12583 | 53092 | 209303 | 840523 | 3358391 | | | | | | | | | | | |
| $\Delta^{12} a_p$ | 40509 | 156211 | 631220 | 2517868 | | | | | | | | | | | | |
| $\Delta^{13} a_p$ | 115702 | 475009 | 1886648 | | | | | | | | | | | | | |
| $\Delta^{14} a_p$ | 359307 | 1411639 | | | | | | | | | | | | | | |
| $\Delta^{15} a_p$ | 1052332 | | | | | | | | | | | | | | | |

### TABLE (6). $0^{th}$ THROUGH $15^{th}$ DIFFERENCES OF $c_p$ FOR $0 \le p \le 15$

| $p$ | 0 | 1 | 2 | 3 | 4 | 5 | 6 | 7 | 8 | 9 | 10 | 11 | 12 | 13 | 14 | 15 |
|---|---|---|---|---|---|---|---|---|---|---|---|---|---|---|---|---|
| $c_p$ | **1** | **1** | **2** | **3** | **5** | **7** | **11** | **15** | **22** | **30** | **42** | **56** | **77** | **101** | **135** | **176** |
| $\Delta c_p$ | 0 | 1 | 1 | 2 | 2 | 4 | 4 | 7 | 8 | 12 | 14 | 21 | 24 | 34 | 41 | |
| $\Delta^2 c_p$ | 1 | 0 | 1 | 0 | 2 | 0 | 3 | 1 | 4 | 2 | 7 | 3 | 10 | 7 | | |
| $\Delta^3 c_p$ | −1 | 1 | −1 | 2 | −2 | 3 | −2 | 3 | −2 | 5 | −4 | 7 | −3 | | | |
| $\Delta^4 c_p$ | 2 | −2 | 3 | −4 | 5 | −5 | 5 | −5 | 7 | −9 | 11 | −10 | | | | |
| $\Delta^5 c_p$ | −4 | 5 | −7 | 9 | −10 | 10 | −10 | 12 | −16 | 20 | −21 | | | | | |
| $\Delta^6 c_p$ | 9 | −12 | 16 | −19 | 20 | −20 | 22 | −28 | 36 | −41 | | | | | | |
| $\Delta^7 c_p$ | −21 | 28 | −35 | 39 | −40 | 42 | −50 | 64 | −77 | | | | | | | |
| $\Delta^8 c_p$ | 49 | −63 | 74 | −79 | 82 | −92 | 114 | −141 | | | | | | | | |
| $\Delta^9 c_p$ | −112 | 137 | −153 | 161 | −174 | 206 | −255 | | | | | | | | | |
| $\Delta^{10} c_p$ | 249 | −290 | 314 | −335 | 380 | −461 | | | | | | | | | | |
| $\Delta^{11} c_p$ | −539 | 604 | −649 | 715 | −841 | | | | | | | | | | | |
| $\Delta^{12} c_p$ | 1143 | −1253 | 1364 | −1556 | | | | | | | | | | | | |
| $\Delta^{13} c_p$ | −2396 | 2617 | −2920 | | | | | | | | | | | | | |
| $\Delta^{14} c_p$ | 5013 | −5537 | | | | | | | | | | | | | | |
| $\Delta^{15} c_p$ | −10550 | | | | | | | | | | | | | | | |

### TABLE (7). $0^{th}$ THROUGH $15^{th}$ DIFFERENCES OF $a'_p = a_p - c_p$ FOR $0 \le p \le 15$

| $p$ | 0 | 1 | 2 | 3 | 4 | 5 | 6 | 7 | 8 | 9 | 10 | 11 | 12 | 13 | 14 | 15 |
|---|---|---|---|---|---|---|---|---|---|---|---|---|---|---|---|---|
| $a'_p$ | **0** | **0** | **1** | **5** | **20** | **78** | **307** | **1219** | **4862** | **19428** | **77685** | **310705** | **1242776** | **4971050** | **19884135** | **79536463** |
| $\Delta a'_p$ | 0 | 1 | 4 | 15 | 58 | 229 | 912 | 3643 | 14566 | 58257 | 233020 | 932071 | 3728274 | 14913085 | 59652328 | |
| $\Delta^2 a'_p$ | 1 | 3 | 11 | 43 | 171 | 683 | 2731 | 10923 | 43691 | 174763 | 699051 | 2796203 | 11184811 | 44739243 | | |
| $\Delta^3 a'_p$ | 2 | 8 | 32 | 128 | 512 | 2048 | 8192 | 32768 | 131072 | 524288 | 2097152 | 8388608 | 33554432 | | | |
| $\Delta^4 a'_p$ | 6 | 24 | 96 | 384 | 1536 | 6144 | 24576 | 98304 | 393216 | 1572864 | 6291456 | 25165824 | | | | |
| $\Delta^5 a'_p$ | 18 | 72 | 288 | 1152 | 4608 | 18432 | 73728 | 294912 | 1179648 | 4718592 | 18874368 | | | | | |
| $\Delta^6 a'_p$ | 54 | 216 | 864 | 3456 | 13824 | 55296 | 221184 | 884736 | 3538944 | 14155776 | | | | | | |
| $\Delta^7 a'_p$ | 162 | 648 | 2592 | 10368 | 41472 | 165888 | 663552 | 2654208 | 10616832 | | | | | | | |
| $\Delta^8 a'_p$ | 486 | 1944 | 7776 | 31104 | 124416 | 497664 | 1990656 | 7962624 | | | | | | | | |
| $\Delta^9 a'_p$ | 1458 | 5832 | 23328 | 93312 | 373248 | 1492992 | 5971968 | | | | | | | | | |
| $\Delta^{10} a'_p$ | 4374 | 17496 | 69984 | 279936 | 1119744 | 4478976 | | | | | | | | | | |
| $\Delta^{11} a'_p$ | 13122 | 52488 | 209952 | 839808 | 3359232 | | | | | | | | | | | |
| $\Delta^{12} a'_p$ | 39366 | 157464 | 629856 | 2519424 | | | | | | | | | | | | |
| $\Delta^{13} a'_p$ | 118098 | 472392 | 1889568 | | | | | | | | | | | | | |
| $\Delta^{14} a'_p$ | 354294 | 1417176 | | | | | | | | | | | | | | |
| $\Delta^{15} a'_p$ | 1062882 | | | | | | | | | | | | | | | |

### TABLE (8). $0^{th}$ THROUGH $15^{th}$ DIFFERENCES OF $b_p$ FOR $0 \le p \le 15$

| $p$ | 0 | 1 | 2 | 3 | 4 | 5 | 6 | 7 | 8 | 9 | 10 | 11 | 12 | 13 | 14 | 15 |
|---|---|---|---|---|---|---|---|---|---|---|---|---|---|---|---|---|
| $b_p$ | **1** | **2** | **7** | **26** | **115** | **596** | **3331** | **19100** | **110329** | **638540** | **3697405** | **21411974** | **124001893** | **718129334** | **4158891979** | **24085338398** |
| $\Delta b_p$ | 1 | 5 | 19 | 89 | 481 | 2735 | 15769 | 91229 | 528211 | 3058865 | 17714569 | 102589919 | 594127441 | 3440762645 | 19926446419 | |
| $\Delta^2 b_p$ | 4 | 14 | 70 | 392 | 2254 | 13034 | 75460 | 436982 | 2530654 | 14655704 | 84875350 | 491537522 | 2846635204 | 16485683774 | | |
| $\Delta^3 b_p$ | 10 | 56 | 322 | 1862 | 10780 | 62426 | 361522 | 2093672 | 12125050 | 70219646 | 406662172 | 2355097682 | 13639048570 | | | |
| $\Delta^4 b_p$ | 46 | 266 | 1540 | 8918 | 51646 | 299096 | 1732150 | 10031378 | 58094596 | 336442526 | 1948435510 | 11283950888 | | | | |
| $\Delta^5 b_p$ | 220 | 1274 | 7378 | 42728 | 247450 | 1433054 | 8299228 | 48063218 | 278347930 | 1611992984 | 9335515378 | | | | | |
| $\Delta^6 b_p$ | 1054 | 6104 | 35350 | 204722 | 1185604 | 6866174 | 39763990 | 230284712 | 1333645054 | 7723522394 | | | | | | |
| $\Delta^7 b_p$ | 5050 | 29246 | 169372 | 980882 | 5680570 | 32897816 | 190520722 | 1103360342 | 6389877340 | | | | | | | |
| $\Delta^8 b_p$ | 24196 | 140126 | 811510 | 4699688 | 27217246 | 157622906 | 912839620 | 5286516998 | | | | | | | | |
| $\Delta^9 b_p$ | 115930 | 671384 | 3888178 | 22517558 | 130405660 | 755216714 | 4373677378 | | | | | | | | | |
| $\Delta^{10} b_p$ | 555454 | 3216794 | 18629380 | 107888102 | 624811054 | 3618460664 | | | | | | | | | | |
| $\Delta^{11} b_p$ | 2661340 | 15412586 | 89258722 | 516922952 | 2993649610 | | | | | | | | | | | |
| $\Delta^{12} b_p$ | 12751246 | 73846136 | 427664230 | 2476726658 | | | | | | | | | | | | |
| $\Delta^{13} b_p$ | 61094890 | 353818094 | 2049062428 | | | | | | | | | | | | | |
| $\Delta^{14} b_p$ | 292723204 | 1695244334 | | | | | | | | | | | | | | |
| $\Delta^{15} b_p$ | 1402521130 | | | | | | | | | | | | | | | |

In Table (7) for $a'_p$ it can be seen that

1. $\forall\ p \geq 0$,

$$a'_p \ (\mathrm{mod}\ 2) = \tfrac{1}{2}\,[1 - (-1)^{p\,(p-1)/2}]; \tag{18}$$

2. $\forall\ p \geq 0$,

$$\Delta\, a'_p = 4\,\Delta\, a'_{p-1} - (p-2); \tag{19}$$

3. $\forall\ p \geq 0$,

$$\Delta\, a'_p \ (\mathrm{mod}\ 10) = 2\,p - \tfrac{1}{2}\,[1 - (-1)^p], \tag{20}$$

whence

$$\Delta\, a'_p \ (\mathrm{mod}\ 10) = \begin{cases} 0, & \text{if } p \equiv 0 \ (\mathrm{mod}\ 10) \\ 1, & \text{if } p \equiv 1 \ (\mathrm{mod}\ 10) \\ 4, & \text{if } p \equiv 2 \ (\mathrm{mod}\ 10) \\ 5, & \text{if } p \equiv 3 \ (\mathrm{mod}\ 10) \\ 8, & \text{if } p \equiv 4 \ (\mathrm{mod}\ 10) \\ 9, & \text{if } p \equiv 5 \ (\mathrm{mod}\ 10) \\ 2, & \text{if } p \equiv 6 \ (\mathrm{mod}\ 10) \\ 3, & \text{if } p \equiv 7 \ (\mathrm{mod}\ 10) \\ 6, & \text{if } p \equiv 8 \ (\mathrm{mod}\ 10) \\ 7, & \text{if } p \equiv 9 \ (\mathrm{mod}\ 10) \end{cases}; \tag{21}$$

4. $\forall\ p \geq 0$,

$$\Delta^2\, a'_p = 3\,\Delta\, a'_p - (p-1); \tag{22}$$

5. $\forall\ p \geq 0$,

$$\Delta^2\, a'_p = 4\,\Delta^2\, a'_{p-1} - 1; \tag{23}$$

6. $\forall\ p \geq 0$,

$$\Delta^2\, a'_p \ (\mathrm{mod}\ 10) = \begin{cases} 1, & \text{if } 2p \equiv 0 \ (\mathrm{mod}\ 4) \\ 3, & \text{if } 2p \equiv 1 \ (\mathrm{mod}\ 4) \end{cases}; \tag{24}$$

7. $\forall\ p \geq 0$,

$$\Delta^3\, a'_p = 3\,\Delta^2\, a'_p - 1; \tag{25}$$

8. $\forall\ k \geq 4$ and $p \geq 0$,

$$\Delta^k\, a'_p = 3\,\Delta^{k-1}\, a'_p, \tag{26}$$

whence, $\forall\ q \geq 0$,

$$\Delta^{k+q}\, a'_p = 3^{q+1}\,\Delta^{k-1}\, a'_p; \tag{27}$$

9. $\forall\ k \geq 3$ and $p \geq 0$,

$$\Delta^k\, a'_p = 4\,\Delta^k\, a'_{p-1}, \tag{28}$$

whence, $\forall\ q \geq 0$,

$$\Delta^k\, a'_{p+q} = 4^{q+1}\,\Delta^k\, a'_{p-1}; \tag{29}$$

and

10. $\forall\ k \geq 3$ and $p \geq 0$,

$$\Delta^k\, a'_p \ (\mathrm{mod}\ 10) = \begin{cases} 6, & \text{if } k + 2p \equiv 0 \ (\mathrm{mod}\ 4) \\ 8, & \text{if } k + 2p \equiv 1 \ (\mathrm{mod}\ 4) \\ 4, & \text{if } k + 2p \equiv 2 \ (\mathrm{mod}\ 4) \\ 2, & \text{if } k + 2p \equiv 3 \ (\mathrm{mod}\ 4) \end{cases}. \tag{30}$$

In Table (8) for $b_p$ it can be seen that

1. $\forall\ p \geq 0$,

$$b_p \ (\mathrm{mod}\ 2) = \tfrac{1}{2}\,[1 - (-1)^{p+1}]; \tag{31}$$

2. $\forall\ p \geq 0$,

$$\Delta\, b_p \ (\mathrm{mod}\ 10) = \begin{cases} 1, & \text{if } p \equiv 0 \ (\mathrm{mod}\ 4) \\ 5, & \text{if } p \equiv 1 \ (\mathrm{mod}\ 4) \\ 9, & \text{if } p \equiv 2 \ (\mathrm{mod}\ 4) \\ 9, & \text{if } p \equiv 3 \ (\mathrm{mod}\ 4) \end{cases}; \tag{32}$$

3. $\forall\ k \equiv 2 \ (\mathrm{mod}\ 4)$ and $p \geq 0$,

$$\Delta^k\, b_p \ (\mathrm{mod}\ 10) = \begin{cases} 4, & \text{if } p \equiv 0 \ (\mathrm{mod}\ 4) \\ 4, & \text{if } p \equiv 1 \ (\mathrm{mod}\ 4) \\ 0, & \text{if } p \equiv 2 \ (\mathrm{mod}\ 4) \\ 2, & \text{if } p \equiv 3 \ (\mathrm{mod}\ 4) \end{cases}; \tag{33}$$

4. $\forall\ k \equiv 3 \ (\mathrm{mod}\ 4)$ and $p \geq 0$,

$$\Delta^k\, b_p \ (\mathrm{mod}\ 10) = \begin{cases} 0, & \text{if } p \equiv 0 \ (\mathrm{mod}\ 4) \\ 6, & \text{if } p \equiv 1 \ (\mathrm{mod}\ 4) \\ 2, & \text{if } p \equiv 2 \ (\mathrm{mod}\ 4) \\ 2, & \text{if } p \equiv 3 \ (\mathrm{mod}\ 4) \end{cases}; \tag{34}$$

5. $\forall\ 0 < k \equiv 0 \ (\mathrm{mod}\ 4)$ and $p \geq 0$,

$$\Delta^k\, b_p \ (\mathrm{mod}\ 10) = \begin{cases} 6, & \text{if } p \equiv 0 \ (\mathrm{mod}\ 4) \\ 6, & \text{if } p \equiv 1 \ (\mathrm{mod}\ 4) \\ 0, & \text{if } p \equiv 2 \ (\mathrm{mod}\ 4) \\ 8, & \text{if } p \equiv 3 \ (\mathrm{mod}\ 4) \end{cases}; \tag{35}$$

6. $\forall\ 1 < k \equiv 1 \ (\mathrm{mod}\ 4)$ and $p \geq 0$,

$$\Delta^k\, b_p \ (\mathrm{mod}\ 10) = \begin{cases} 0\ (= 1 - 1), & \text{if } p \equiv 0 \ (\mathrm{mod}\ 4) \\ 4\ (= 5 - 1), & \text{if } p \equiv 1 \ (\mathrm{mod}\ 4) \\ 8\ (= 9 - 1), & \text{if } p \equiv 2 \ (\mathrm{mod}\ 4) \\ 8\ (= 9 - 1), & \text{if } p \equiv 3 \ (\mathrm{mod}\ 4) \end{cases}; \tag{36}$$

and

7. $\forall\ k \geq 0$ and $p \geq 0$,

$$\Delta^{k+2}\, b_{p+2} = 7\,\Delta^{k+3}\, b_p, \tag{37}$$

whence, $\forall\ 2q \leq p$,

$$\Delta^{k+2}\, b_{p+2} = 7^{q+1}\,\Delta^{k+q+3}\, b_{p-2q}. \tag{38}$$

Values for the numbers $a_p$ and $b_p$—together with some other features of the general expressions for $L_{(p)}$ and $G_{(p)a}{}^b$, such as the numbers of covariant index permutations per Eqs. (1) and (2) comprehended by, and the overall sums of the magnitudes of the numerical coefficients of the terms appearing in, the expressions in question—appear below in Tables (9) and (10).

The $1^{st}$ $a_{p-1}$ terms of general expressions for $L_{(p)}$ can be calculated directly (for $p \geq 1$) by means of the inverse of the formula

$$\frac{\partial}{\partial R}\, L_{(p)} = -\,p\, L_{(p-1)}. \tag{39}$$

The initial terms (up to 10) of expressions for $L_{(p)}$ for $0 \leq p \leq 15$ calculated by means—as needs be—of the inverse of Eq. (39) appear below in Eqs. (40) through (55).

$$L_{(0)} = +\,1; \tag{40}$$

$$L_{(1)} = -\,R; \tag{41}$$

$$L_{(2)} = R^2 - 4\, R_b{}^a\, R_a{}^b + R_{cd}{}^{ab}\, R_{ab}{}^{cd}; \tag{42}$$

$$\begin{aligned} L_{(3)} = {} & -R^3 + 12\, R\, R_b{}^a\, R_a{}^b - 3\, R\, R_{cd}{}^{ab}\, R_{ab}{}^{cd} - 16\, R_b{}^a\, R_c{}^b\, R_a{}^c + \\ & + 24\, R_c{}^a\, R_d{}^b\, R_{ab}{}^{cd} + 24\, R_b{}^a\, R_{de}{}^{bc}\, R_{ac}{}^{de} + 2\, R_{cd}{}^{ab}\, R_{ef}{}^{cd}\, R_{ab}{}^{ef} - \\ & - 8\, R_{ce}{}^{ab}\, R_{af}{}^{cd}\, R_{bd}{}^{ef}; \end{aligned} \tag{43}$$

$$\begin{aligned} L_{(4)} = {} & R^4 - 24\, R^2\, R_a{}^b\, R_b{}^a + 6\, R^2\, R_{ab}{}^{cd}\, R_{cd}{}^{ab} + 64\, R\, R_a{}^c\, R_b{}^a\, R_c{}^b - \\ & - 96\, R\, R_c{}^a\, R_d{}^b\, R_{ab}{}^{cd} - 96\, R\, R_b{}^a\, R_{ac}{}^{de}\, R_{de}{}^{bc} - \\ & - 8\, R\, R_{ab}{}^{ef}\, R_{cd}{}^{ab}\, R_{ef}{}^{cd} + 32\, R\, R_{af}{}^{cd}\, R_{bd}{}^{ef}\, R_{ce}{}^{ab} + \\ & + 48\, R_a{}^b\, R_b{}^a\, R_c{}^d\, R_d{}^c - 96\, R_a{}^d\, R_b{}^a\, R_c{}^b\, R_d{}^c \ \ldots, \end{aligned} \tag{44}$$

there being 15 additional terms, the $1^{st}$ of which is positive;

$$\begin{aligned} L_{(5)} = {} & -R^5 + 40\, R^3\, R_a{}^b\, R_b{}^a - 10\, R^3\, R_{ab}{}^{cd}\, R_{cd}{}^{ab} - 160\, R^2\, R_a{}^c\, R_b{}^a\, R_c{}^b + \\ & + 240\, R^2\, R_c{}^a\, R_d{}^b\, R_{ab}{}^{cd} + 240\, R^2\, R_b{}^a\, R_{ac}{}^{de}\, R_{de}{}^{bc} + \\ & + 20\, R^2\, R_{ab}{}^{ef}\, R_{cd}{}^{ab}\, R_{ef}{}^{cd} - 80\, R^2\, R_{af}{}^{cd}\, R_{bd}{}^{ef}\, R_{ce}{}^{ab} - \\ & - 240\, R\, R_a{}^b\, R_b{}^a\, R_c{}^d\, R_d{}^c + 480\, R\, R_a{}^d\, R_b{}^a\, R_c{}^b\, R_d{}^c \ \ldots, \end{aligned} \tag{45}$$

there being 75 additional terms, the $1^{st}$ of which is negative;

$$\begin{aligned} L_{(6)} = {} & R^6 - 60\, R^4\, R_a{}^b\, R_b{}^a + 15\, R^4\, R_{ab}{}^{cd}\, R_{cd}{}^{ab} + 320\, R^3\, R_a{}^c\, R_b{}^a\, R_c{}^b - \\ & - 480\, R^3\, R_c{}^a\, R_d{}^b\, R_{ab}{}^{cd} - 480\, R^3\, R_b{}^a\, R_{ac}{}^{de}\, R_{de}{}^{bc} - \\ & - 40\, R^3\, R_{ab}{}^{ef}\, R_{cd}{}^{ab}\, R_{ef}{}^{cd} + 160\, R^3\, R_{af}{}^{cd}\, R_{bd}{}^{ef}\, R_{ce}{}^{ab} + \\ & + 720\, R^2\, R_a{}^b\, R_b{}^a\, R_c{}^d\, R_d{}^c - 1440\, R^2\, R_a{}^d\, R_b{}^a\, R_c{}^b\, R_d{}^c \ \ldots, \end{aligned} \tag{46}$$

TABLE (9). SOME PROPERTIES OF $L_{(p)}$ FOR $0 \le p \le 15$

| ORDER | CURVATURE DEPENDENCE | QUANTITY | NUMBER OF TERMS | NUMBER OF PERMUTATIONS COMPREHENDED | OVERALL SUM OF THE MAGNITUDES OF THE NUMERICAL FACTORS |
|---|---|---|---|---|---|
| $p$ | — | $L_{(p)}$ | $a_p$ | $(2p)!$ | $\frac{(2p)!}{2^p}$ |
| 0 | Zero | $L_{(0)}$ | **1** | 1 | 1 |
| 1 | Linear | $L_{(1)}$ | **1** | 2 | 1 |
| 2 | Quadratic | $L_{(2)}$ | **3** | 24 | 6 |
| 3 | Cubic | $L_{(3)}$ | **8** | 720 | 90 |
| 4 | Quartic | $L_{(4)}$ | **25** | 40,320 | 2520 |
| 5 | Quintic | $L_{(5)}$ | **85** | 3,628,800 | 113,400 |
| 6 | Sextic | $L_{(6)}$ | **318** | 479,001,600 | 7,484,400 |
| 7 | Septic | $L_{(7)}$ | **1234** | 87,178,291,200 | 681,080,400 |
| 8 | Octic | $L_{(8)}$ | **4884** | 20,922,789,888,000 | 81,729,648,000 |
| 9 | Nonic | $L_{(9)}$ | **19,458** | 6,402,373,705,728,000 | 12,504,636,144,000 |
| 10 | Decic | $L_{(10)}$ | **77,727** | 2,432,902,008,176,640,000 | 2,375,880,867,360,000 |
| 11 | Undecic | $L_{(11)}$ | **310,761** | 1,124,000,727,777,607,680,000 | 548,828,480,360,160,000 |
| 12 | Duodecic | $L_{(12)}$ | **1,242,853** | 620,448,401,733,239,439,360,000 | 151,476,660,579,404,160,000 |
| 13 | Tredecic | $L_{(13)}$ | **4,971,151** | 403,291,461,126,605,635,584,000,000 | 49,229,914,688,306,352,000,000 |
| 14 | Quattuordecic | $L_{(14)}$ | **19,884,270** | 304,888,344,611,713,860,501,504,000,000 | 18,608,907,752,179,801,056,000,000 |
| 15 | Quindecic | $L_{(15)}$ | **79,536,639** | 265,252,859,812,191,058,636,308,480,000,000 | 8,094,874,872,198,213,459,360,000,000 |

TABLE (10). SOME PROPERTIES OF ${G_{(p)a}}^b$ FOR $0 \le p \le 15$

| ORDER | CURVATURE DEPENDENCE | QUANTITY | NUMBER OF TERMS | NUMBER OF PERMUTATIONS COMPREHENDED | OVERALL SUM OF THE MAGNITUDES OF THE NUMERICAL FACTORS |
|---|---|---|---|---|---|
| $p$ | — | ${G_{(p)a}}^b$ | $b_p$ | $(2p+1)!$ | 1 if $p = 0$; $\frac{(2p+1)!}{2^{p+1}p}$ if $p \ge 1$ |
| 0 | Zero | ${G_{(0)a}}^b$ | **1** | 1 | 1 |
| 1 | Linear | ${G_{(1)a}}^b$ | **2** | 6 | 1 1/2 |
| 2 | Quadratic | ${G_{(2)a}}^b$ | **7** | 120 | 7 1/2 |
| 3 | Cubic | ${G_{(3)a}}^b$ | **26** | 5040 | 105 |
| 4 | Quartic | ${G_{(4)a}}^b$ | **115** | 362,880 | 2835 |
| 5 | Quintic | ${G_{(5)a}}^b$ | **596** | 39,916,800 | 124,740 |
| 6 | Sextic | ${G_{(6)a}}^b$ | **3331** | 6,227,020,800 | 8,108,100 |
| 7 | Septic | ${G_{(7)a}}^b$ | **19,100** | 1,307,674,368,000 | 729,729,000 |
| 8 | Octic | ${G_{(8)a}}^b$ | **110,329** | 355,687,428,096,000 | 86,837,751,000 |
| 9 | Nonic | ${G_{(9)a}}^b$ | **638,540** | 121,645,100,408,832,000 | 13,199,338,152,000 |
| 10 | Decic | ${G_{(10)a}}^b$ | **3,697,405** | 51,090,942,171,709,440,000 | 2,494,674,910,728,000 |
| 11 | Undecic | ${G_{(11)a}}^b$ | **21,411,974** | 25,852,016,738,884,976,640,000 | 573,775,229,467,440,000 |
| 12 | Duodecic | ${G_{(12)a}}^b$ | **124,001,893** | 15,511,210,043,330,985,984,000,000 | 157,788,188,103,546,000,000 |
| 13 | Tredecic | ${G_{(13)a}}^b$ | **718,129,334** | 10,888,869,450,418,352,160,768,000,000 | 51,123,372,945,548,904,000,000 |
| 14 | Quattuordecic | ${G_{(14)a}}^b$ | **4,158,891,979** | 8,841,761,993,739,701,954,543,616,000,000 | 19,273,511,600,471,936,808,000,000 |
| 15 | Quindecic | ${G_{(15)a}}^b$ | **24,085,338,398** | 8,222,838,654,177,922,817,725,562,880,000,000 | 8,364,704,034,604,820,574,672,000,000 |

there being 308 additional terms, the 1st of which is positive;

$$L_{(7)} = -R^7 + 84\,R^5 R_a{}^b R_b{}^a - 21\,R^5 R_{ab}{}^{cd} R_{cd}{}^{ab} - 560\,R^4 R_a{}^c R_b{}^a R_c{}^b + 840\,R^4 R_c{}^a R_d{}^b R_{ab}{}^{cd} + 840\,R^4 R_b{}^a R_{ac}{}^{de} R_{de}{}^{bc} + 70\,R^4 R_{ab}{}^{ef} R_{cd}{}^{ab} R_{ef}{}^{cd} - \\ - 280\,R^4 R_{af}{}^{cd} R_{bd}{}^{ef} R_{ce}{}^{ab} - 1680\,R^3 R_a{}^b R_b{}^a R_c{}^d R_d{}^c + 3360\,R^3 R_a{}^d R_b{}^a R_c{}^b R_d{}^c \ldots, \tag{47}$$

there being 1224 additional terms, the 1st of which is negative;

$$L_{(8)} = R^8 - 112\,R^6 R_a{}^b R_b{}^a + 28\,R^6 R_{ab}{}^{cd} R_{cd}{}^{ab} + 896\,R^5 R_a{}^c R_b{}^a R_c{}^b - 1344\,R^5 R_c{}^a R_d{}^b R_{ab}{}^{cd} - 1344\,R^5 R_b{}^a R_{ac}{}^{de} R_{de}{}^{bc} - 112\,R^5 R_{ab}{}^{ef} R_{cd}{}^{ab} R_{ef}{}^{cd} + \\ + 448\,R^5 R_{af}{}^{cd} R_{bd}{}^{ef} R_{ce}{}^{ab} + 3360\,R^4 R_a{}^b R_b{}^a R_c{}^d R_d{}^c - 6720\,R^4 R_a{}^d R_b{}^a R_c{}^b R_d{}^c \ldots, \tag{48}$$

there being 4874 additional terms, the 1st of which is positive;

$$L_{(9)} = -R^9 + 144\,R^7 R_a{}^b R_b{}^a - 36\,R^7 R_{ab}{}^{cd} R_{cd}{}^{ab} - 1344\,R^6 R_a{}^c R_b{}^a R_c{}^b + 2016\,R^6 R_c{}^a R_d{}^b R_{ab}{}^{cd} + 2016\,R^6 R_b{}^a R_{ac}{}^{de} R_{de}{}^{bc} + \\ + 168\,R^6 R_{ab}{}^{ef} R_{cd}{}^{ab} R_{ef}{}^{cd} - 672\,R^6 R_{af}{}^{cd} R_{bd}{}^{ef} R_{ce}{}^{ab} - 6048\,R^5 R_a{}^b R_b{}^a R_c{}^d R_d{}^c + 12{,}096\,R^5 R_a{}^d R_b{}^a R_c{}^b R_d{}^c \ldots, \tag{49}$$

there being 19,448 additional terms, the 1st of which is negative;

$$L_{(10)} = R^{10} - 180\,R^8 R_a{}^b R_b{}^a + 45\,R^8 R_{ab}{}^{cd} R_{cd}{}^{ab} + 1920\,R^7 R_a{}^c R_b{}^a R_c{}^b - 2880\,R^7 R_c{}^a R_d{}^b R_{ab}{}^{cd} - 2880\,R^7 R_b{}^a R_{ac}{}^{de} R_{de}{}^{bc} - \\ - 240\,R^7 R_{ab}{}^{ef} R_{cd}{}^{ab} R_{ef}{}^{cd} + 960\,R^7 R_{af}{}^{cd} R_{bd}{}^{ef} R_{ce}{}^{ab} + 10{,}080\,R^6 R_a{}^b R_b{}^a R_c{}^d R_d{}^c - 20{,}160\,R^6 R_a{}^d R_b{}^a R_c{}^b R_d{}^c \ldots, \tag{50}$$

there being 77,717 additional terms, the 1st of which is positive;

$$L_{(11)} = -R^{11} + 220\,R^9 R_a{}^b R_b{}^a - 55\,R^9 R_{ab}{}^{cd} R_{cd}{}^{ab} - 2640\,R^8 R_a{}^c R_b{}^a R_c{}^b + 3960\,R^8 R_c{}^a R_d{}^b R_{ab}{}^{cd} + 3960\,R^8 R_b{}^a R_{ac}{}^{de} R_{de}{}^{bc} + \\ + 330\,R^8 R_{ab}{}^{ef} R_{cd}{}^{ab} R_{ef}{}^{cd} - 1320\,R^8 R_{af}{}^{cd} R_{bd}{}^{ef} R_{ce}{}^{ab} - 15{,}840\,R^7 R_a{}^b R_b{}^a R_c{}^d R_d{}^c + 31{,}680\,R^7 R_a{}^d R_b{}^a R_c{}^b R_d{}^c \ldots, \tag{51}$$

there being 310,751 additional terms, the 1st of which is negative;

$$L_{(12)} = R^{12} - 264\,R^{10} R_a{}^b R_b{}^a + 66\,R^{10} R_{ab}{}^{cd} R_{cd}{}^{ab} + 3520\,R^9 R_a{}^c R_b{}^a R_c{}^b - 5280\,R^9 R_c{}^a R_d{}^b R_{ab}{}^{cd} - 5280\,R^9 R_b{}^a R_{ac}{}^{de} R_{de}{}^{bc} - \\ - 440\,R^9 R_{ab}{}^{ef} R_{cd}{}^{ab} R_{ef}{}^{cd} + 1760\,R^9 R_{af}{}^{cd} R_{bd}{}^{ef} R_{ce}{}^{ab} + 23{,}760\,R^8 R_a{}^b R_b{}^a R_c{}^d R_d{}^c - 47{,}520\,R^8 R_a{}^d R_b{}^a R_c{}^b R_d{}^c \ldots, \tag{52}$$

there being 1,242,843 additional terms, the 1st of which is positive;

$$L_{(13)} = -R^{13} + 312\,R^{11} R_a{}^b R_b{}^a - 78\,R^{11} R_{ab}{}^{cd} R_{cd}{}^{ab} - 4576\,R^{10} R_a{}^c R_b{}^a R_c{}^b + 6864\,R^{10} R_c{}^a R_d{}^b R_{ab}{}^{cd} + 6864\,R^{10} R_b{}^a R_{ac}{}^{de} R_{de}{}^{bc} + \\ + 572\,R^{10} R_{ab}{}^{ef} R_{cd}{}^{ab} R_{ef}{}^{cd} - 2288\,R^{10} R_{af}{}^{cd} R_{bd}{}^{ef} R_{ce}{}^{ab} - 34{,}320\,R^9 R_a{}^b R_b{}^a R_c{}^d R_d{}^c + 68{,}640\,R^9 R_a{}^d R_b{}^a R_c{}^b R_d{}^c \ldots, \tag{53}$$

there being 4,971,141 additional terms, the 1st of which is negative;

$$L_{(14)} = R^{14} - 364\,R^{12} R_a{}^b R_b{}^a + 91\,R^{12} R_{ab}{}^{cd} R_{cd}{}^{ab} + 5824\,R^{11} R_a{}^c R_b{}^a R_c{}^b - 8736\,R^{11} R_c{}^a R_d{}^b R_{ab}{}^{cd} - 8736\,R^{11} R_b{}^a R_{ac}{}^{de} R_{de}{}^{bc} - \\ - 728\,R^{11} R_{ab}{}^{ef} R_{cd}{}^{ab} R_{ef}{}^{cd} + 2912\,R^{11} R_{af}{}^{cd} R_{bd}{}^{ef} R_{ce}{}^{ab} + 48{,}048\,R^{10} R_a{}^b R_b{}^a R_c{}^d R_d{}^c - 96{,}096\,R^{10} R_a{}^d R_b{}^a R_c{}^b R_d{}^c \ldots, \tag{54}$$

there being 19,884,260 additional terms, the 1st of which is positive; and

$$L_{(15)} = -R^{15} + 420\,R^{13} R_a{}^b R_b{}^a - 105\,R^{13} R_{ab}{}^{cd} R_{cd}{}^{ab} - 7280\,R^{12} R_a{}^c R_b{}^a R_c{}^b + 10{,}920\,R^{12} R_c{}^a R_d{}^b R_{ab}{}^{cd} + 10{,}920\,R^{12} R_b{}^a R_{ac}{}^{de} R_{de}{}^{bc} + \\ + 910\,R^{12} R_{ab}{}^{ef} R_{cd}{}^{ab} R_{ef}{}^{cd} - 3640\,R^{12} R_{af}{}^{cd} R_{bd}{}^{ef} R_{ce}{}^{ab} - 65{,}520\,R^{11} R_a{}^b R_b{}^a R_c{}^d R_d{}^c + 131{,}040\,R^{11} R_a{}^d R_b{}^a R_c{}^b R_d{}^c \ldots, \tag{55}$$

there being 79,536,629 additional terms, the 1st of which is negative.

In general, the initial terms (up to 10) of expressions for $L_{(p)}$ are given by

$$L_{(p)} = (-1)^p R^p + 4\,(-1)^{p+1} \binom{p}{p-2} R^{p-2} R_a{}^b R_b{}^a + (-1)^p \binom{p}{p-2} R^{p-2} R_{ab}{}^{cd} R_{cd}{}^{ab} + 16\,(-1)^p \binom{p}{p-3} R^{p-3} R_a{}^c R_b{}^a R_c{}^b + \\ + 24\,(-1)^{p+1} \binom{p}{p-3} R^{p-3} R_c{}^a R_d{}^b R_{ab}{}^{cd} + 24\,(-1)^{p+1} \binom{p}{p-3} R^{p-3} R_b{}^a R_{ac}{}^{de} R_{de}{}^{bc} + 2\,(-1)^{p+1} \binom{p}{p-3} R^{p-3} R_{ab}{}^{ef} R_{cd}{}^{ab} R_{ef}{}^{cd} + \\ + 8\,(-1)^p \binom{p}{p-3} R^{p-3} R_{af}{}^{cd} R_{bd}{}^{ef} R_{ce}{}^{ab} + 48\,(-1)^p \binom{p}{p-4} R^{p-4} R_a{}^b R_b{}^a R_c{}^d R_d{}^c + 96\,(-1)^{p+1} \binom{p}{p-4} R^{p-4} R_a{}^d R_b{}^a R_c{}^b R_d{}^c \ldots, \tag{56}$$

there being $a_p - 10$ additional terms, the sign of the 1st of which equals $(-1)^p$.

The 1st $a_{p-1}$ terms (in addition to $b_{p-1} - a_{p-1}$ of the last $b_p - a_p$ of the remaining $b_p - a_{p-1}$ terms) of general expressions for $G_{(p)a}{}^b$ can be calculated directly (for $p \geq 2$) by means of the inverse of the formula

$$\frac{\partial}{\partial R}\, G_{(p)a}{}^b = -(p-1)\, G_{(p-1)a}{}^b. \tag{57}$$

The initial terms (up to 10) of expressions for $G_{(p)a}{}^b$ for $0 \leq p \leq 15$ calculated by means—as needs be—of the inverse of Eq. (57) appear below in Eqs. (58) through (73).

$$G_{(0)a}{}^b = \delta_a^b; \tag{58}$$

$$G_{(1)a}{}^b = \frac{1}{2}\left(-\delta_a^b R + 2\,R_a{}^b\right); \tag{59}$$

$$G_{(2)a}{}^b = \frac{1}{4}\left(\delta_a^b R^2 - 4\,\delta_a^b R_d{}^c R_c{}^d + \delta_a^b R_{ef}{}^{cd} R_{cd}{}^{ef} - 4\,R_a{}^b R + 8\,R_a{}^c R_c{}^b - 8\,R_{ad}{}^{bc} R_c{}^d - 4\,R_{ae}{}^{cd} R_{cd}{}^{be}\right); \tag{60}$$

$$G_{(3)a}{}^b = \frac{1}{6}\Big(-\delta_a^b R^3 + 12\,\delta_a^b R\,R_d{}^c R_c{}^d - 3\,\delta_a^b R\,R_{ef}{}^{cd} R_{cd}{}^{ef} - 16\,\delta_a^b R_d{}^c R_e{}^d R_c{}^e + 24\,\delta_a^b R_e{}^c R_f{}^d R_{cd}{}^{ef} + 24\,\delta_a^b R_d{}^c R_{fg}{}^{de} R_{ce}{}^{fg} + 2\,\delta_a^b R_{ef}{}^{cd} R_{gh}{}^{ef} R_{cd}{}^{gh} - \\ - 8\,\delta_a^b R_{eg}{}^{cd} R_{ch}{}^{ef} R_{df}{}^{gh} + 6\,R_a{}^b R^2 - 24\,R_a{}^c R\,R_c{}^b \ldots\Big), \tag{61}$$

there being 16 additional terms, the 1st of which is negative;

$$G_{(4)a}{}^b = \frac{1}{8}\Big(\delta_b^a R^4 - 24\,\delta_b^a R^2 R_c{}^d R_d{}^c + 6\,\delta_b^a R^2 R_{cd}{}^{ef} R_{ef}{}^{cd} + 64\,\delta_b^a R\,R_c{}^e R_d{}^c R_e{}^d - 96\,\delta_b^a R\,R_e{}^c R_f{}^d R_{cd}{}^{ef} - 96\,\delta_b^a R\,R_d{}^c R_{ce}{}^{fg} R_{fg}{}^{de} -$$

$$- 8\, \delta^a_b R\, R_{cd}{}^{gh} R_{ef}{}^{cd} R_{gh}{}^{ef} + 32\, \delta^a_b R\, R_{ch}{}^{ef} R_{df}{}^{gh} R_{eg}{}^{cd} + 48\, \delta^a_b R_c{}^d R_d{}^c R_e{}^f R_f{}^e - 96\, \delta^a_b R_c{}^f R_d{}^c R_e{}^d R_f{}^e \ldots), \qquad (62)$$

there being 105 additional terms, the $1^{st}$ of which is positive;

$$G_{(5)a}{}^b = \tfrac{1}{10} (- \delta^a_b R^5 + 40\, \delta^a_b R^3 R_c{}^d R_d{}^c - 10\, \delta^a_b R^3 R_{cd}{}^{ef} R_{ef}{}^{cd} - 160\, \delta^a_b R^2 R_c{}^e R_d{}^c R_e{}^d + 240\, \delta^a_b R^2 R_e{}^c R_f{}^d R_{cd}{}^{ef} + 240\, \delta^a_b R^2 R_d{}^c R_{ce}{}^{fg} R_{fg}{}^{de} + 20\, \delta^a_b R^2 R_{cd}{}^{gh} R_{ef}{}^{cd} R_{gh}{}^{ef} - 80\, \delta^a_b R^2 R_{ch}{}^{ef} R_{df}{}^{gh} R_{eg}{}^{cd} - 240\, \delta^a_b R\, R_c{}^d R_d{}^c R_e{}^f R_f{}^e + 480\, \delta^a_b R\, R_c{}^f R_d{}^c R_e{}^d R_f{}^e \ldots), \qquad (63)$$

there being 586 additional terms, the $1^{st}$ of which is negative;

$$G_{(6)a}{}^b = \tfrac{1}{12} (\delta^a_b R^6 - 60\, \delta^a_b R^4 R_c{}^d R_d{}^c + 15\, \delta^a_b R^4 R_{cd}{}^{ef} R_{ef}{}^{cd} + 320\, \delta^a_b R^3 R_c{}^e R_d{}^c R_e{}^d - 480\, \delta^a_b R^3 R_e{}^c R_f{}^d R_{cd}{}^{ef} - 480\, \delta^a_b R^3 R_d{}^c R_{ce}{}^{fg} R_{fg}{}^{de} - 40\, \delta^a_b R^3 R_{cd}{}^{gh} R_{ef}{}^{cd} R_{gh}{}^{ef} + 160\, \delta^a_b R^3 R_{ch}{}^{ef} R_{df}{}^{gh} R_{eg}{}^{cd} + 720\, \delta^a_b R^2 R_c{}^d R_d{}^c R_e{}^f R_f{}^e - 1440\, \delta^a_b R^2 R_c{}^f R_d{}^c R_e{}^d R_f{}^e \ldots), \qquad (64)$$

there being 3321 additional terms, the $1^{st}$ of which is positive;

$$G_{(7)a}{}^b = \tfrac{1}{14} (- \delta^a_b R^7 + 84\, \delta^a_b R^5 R_c{}^d R_d{}^c - 21\, \delta^a_b R^5 R_{cd}{}^{ef} R_{ef}{}^{cd} - 560\, \delta^a_b R^4 R_c{}^e R_d{}^c R_e{}^d + 840\, \delta^a_b R^4 R_e{}^c R_f{}^d R_{cd}{}^{ef} + 840\, \delta^a_b R^4 R_d{}^c R_{ce}{}^{fg} R_{fg}{}^{de} + 70\, \delta^a_b R^4 R_{cd}{}^{gh} R_{ef}{}^{cd} R_{gh}{}^{ef} - 280\, \delta^a_b R^4 R_{ch}{}^{ef} R_{df}{}^{gh} R_{eg}{}^{cd} - 1680\, \delta^a_b R^3 R_c{}^d R_d{}^c R_e{}^f R_f{}^e + 3360\, \delta^a_b R^3 R_c{}^f R_d{}^c R_e{}^d R_f{}^e \ldots), \qquad (65)$$

there being 19,090 additional terms, the $1^{st}$ of which is negative;

$$G_{(8)a}{}^b = \tfrac{1}{16} (\delta^a_b R^8 - 112\, \delta^a_b R^6 R_c{}^d R_d{}^c + 28\, \delta^a_b R^6 R_{cd}{}^{ef} R_{ef}{}^{cd} + 896\, \delta^a_b R^5 R_c{}^e R_d{}^c R_e{}^d - 1344\, \delta^a_b R^5 R_e{}^c R_f{}^d R_{cd}{}^{ef} - 1344\, \delta^a_b R^5 R_d{}^c R_{ce}{}^{fg} R_{fg}{}^{de} - 112\, \delta^a_b R^5 R_{cd}{}^{gh} R_{ef}{}^{cd} R_{gh}{}^{ef} + 448\, \delta^a_b R^5 R_{ch}{}^{ef} R_{df}{}^{gh} R_{eg}{}^{cd} + 3360\, \delta^a_b R^4 R_c{}^d R_d{}^c R_e{}^f R_f{}^e - 6720\, \delta^a_b R^4 R_c{}^f R_d{}^c R_e{}^d R_f{}^e \ldots), \qquad (66)$$

there being 110,319 additional terms, the $1^{st}$ of which is positive;

$$G_{(9)a}{}^b = \tfrac{1}{18} (- \delta^a_b R^9 + 144\, \delta^a_b R^7 R_c{}^d R_d{}^c - 36\, \delta^a_b R^7 R_{cd}{}^{ef} R_{ef}{}^{cd} - 1344\, \delta^a_b R^6 R_c{}^e R_d{}^c R_e{}^d + 2016\, \delta^a_b R^6 R_e{}^c R_f{}^d R_{cd}{}^{ef} + 2016\, \delta^a_b R^6 R_d{}^c R_{ce}{}^{fg} R_{fg}{}^{de} + 168\, \delta^a_b R^6 R_{cd}{}^{gh} R_{ef}{}^{cd} R_{gh}{}^{ef} - 672\, \delta^a_b R^6 R_{ch}{}^{ef} R_{df}{}^{gh} R_{eg}{}^{cd} - 6048\, \delta^a_b R^5 R_c{}^d R_d{}^c R_e{}^f R_f{}^e + 12{,}096\, \delta^a_b R^5 R_c{}^f R_d{}^c R_e{}^d R_f{}^e \ldots), \qquad (67)$$

there being 638,530 additional terms, the $1^{st}$ of which is negative;

$$G_{(10)a}{}^b = \tfrac{1}{20} (\delta^a_b R^{10} - 180\, \delta^a_b R^8 R_c{}^d R_d{}^c + 45\, \delta^a_b R^8 R_{cd}{}^{ef} R_{ef}{}^{cd} + 1920\, \delta^a_b R^7 R_c{}^e R_d{}^c R_e{}^d - 2880\, \delta^a_b R^7 R_e{}^c R_f{}^d R_{cd}{}^{ef} - 2880\, \delta^a_b R^7 R_d{}^c R_{ce}{}^{fg} R_{fg}{}^{de} - 240\, \delta^a_b R^7 R_{cd}{}^{gh} R_{ef}{}^{cd} R_{gh}{}^{ef} + 960\, \delta^a_b R^7 R_{ch}{}^{ef} R_{df}{}^{gh} R_{eg}{}^{cd} + 10{,}080\, \delta^a_b R^6 R_c{}^d R_d{}^c R_e{}^f R_f{}^e - 20{,}160\, \delta^a_b R^6 R_c{}^f R_d{}^c R_e{}^d R_f{}^e \ldots), \qquad (68)$$

there being 3,697,395 additional terms, the $1^{st}$ of which is positive;

$$G_{(11)a}{}^b = \tfrac{1}{22} (- \delta^a_b R^{11} + 220\, \delta^a_b R^9 R_c{}^d R_d{}^c - 55\, \delta^a_b R^9 R_{cd}{}^{ef} R_{ef}{}^{cd} - 2640\, \delta^a_b R^8 R_c{}^e R_d{}^c R_e{}^d + 3960\, \delta^a_b R^8 R_e{}^c R_f{}^d R_{cd}{}^{ef} + 3960\, \delta^a_b R^8 R_d{}^c R_{ce}{}^{fg} R_{fg}{}^{de} + 330\, \delta^a_b R^8 R_{cd}{}^{gh} R_{ef}{}^{cd} R_{gh}{}^{ef} - 1320\, \delta^a_b R^8 R_{ch}{}^{ef} R_{df}{}^{gh} R_{eg}{}^{cd} - 15{,}840\, \delta^a_b R^7 R_c{}^d R_d{}^c R_e{}^f R_f{}^e + 31{,}680\, \delta^a_b R^7 R_c{}^f R_d{}^c R_e{}^d R_f{}^e \ldots), \qquad (69)$$

there being 21,411,964 additional terms, the $1^{st}$ of which is negative;

$$G_{(12)a}{}^b = \tfrac{1}{24} (\delta^a_b R^{12} - 264\, \delta^a_b R^{10} R_c{}^d R_d{}^c + 66\, \delta^a_b R^{10} R_{cd}{}^{ef} R_{ef}{}^{cd} + 3520\, \delta^a_b R^9 R_c{}^e R_d{}^c R_e{}^d - 5280\, \delta^a_b R^9 R_e{}^c R_f{}^d R_{cd}{}^{ef} - 5280\, \delta^a_b R^9 R_d{}^c R_{ce}{}^{fg} R_{fg}{}^{de} - 440\, \delta^a_b R^9 R_{cd}{}^{gh} R_{ef}{}^{cd} R_{gh}{}^{ef} + 1760\, \delta^a_b R^9 R_{ch}{}^{ef} R_{df}{}^{gh} R_{eg}{}^{cd} + 23{,}760\, \delta^a_b R^8 R_c{}^d R_d{}^c R_e{}^f R_f{}^e - 47{,}520\, \delta^a_b R^8 R_c{}^f R_d{}^c R_e{}^d R_f{}^e \ldots), \qquad (70)$$

there being 124,001,883 additional terms, the $1^{st}$ of which is positive;

$$G_{(13)a}{}^b = \tfrac{1}{26} (- \delta^a_b R^{13} + 312\, \delta^a_b R^{11} R_c{}^d R_d{}^c - 78\, \delta^a_b R^{11} R_{cd}{}^{ef} R_{ef}{}^{cd} - 4576\, \delta^a_b R^{10} R_c{}^e R_d{}^c R_e{}^d + 6864\, \delta^a_b R^{10} R_e{}^c R_f{}^d R_{cd}{}^{ef} + 6864\, \delta^a_b R^{10} R_d{}^c R_{ce}{}^{fg} R_{fg}{}^{de} + 572\, \delta^a_b R^{10} R_{cd}{}^{gh} R_{ef}{}^{cd} R_{gh}{}^{ef} - 2288\, \delta^a_b R^{10} R_{ch}{}^{ef} R_{df}{}^{gh} R_{eg}{}^{cd} - 34{,}320\, \delta^a_b R^9 R_c{}^d R_d{}^c R_e{}^f R_f{}^e + 68{,}640\, \delta^a_b R^9 R_c{}^f R_d{}^c R_e{}^d R_f{}^e \ldots), \qquad (71)$$

there being 718,129,324 additional terms, the $1^{st}$ of which is negative;

$$G_{(14)a}{}^b = \tfrac{1}{28} (\delta^a_b R^{14} - 364\, \delta^a_b R^{12} R_c{}^d R_d{}^c + 91\, \delta^a_b R^{12} R_{cd}{}^{ef} R_{ef}{}^{cd} + 5824\, \delta^a_b R^{11} R_c{}^e R_d{}^c R_e{}^d - 8736\, \delta^a_b R^{11} R_e{}^c R_f{}^d R_{cd}{}^{ef} - 8736\, \delta^a_b R^{11} R_d{}^c R_{ce}{}^{fg} R_{fg}{}^{de} - 728\, \delta^a_b R^{11} R_{cd}{}^{gh} R_{ef}{}^{cd} R_{gh}{}^{ef} + 2912\, \delta^a_b R^{11} R_{ch}{}^{ef} R_{df}{}^{gh} R_{eg}{}^{cd} + 48{,}048\, \delta^a_b R^{10} R_c{}^d R_d{}^c R_e{}^f R_f{}^e - 96{,}096\, \delta^a_b R^{10} R_c{}^f R_d{}^c R_e{}^d R_f{}^e \ldots), \qquad (72)$$

there being 4,158,891,969 additional terms, the $1^{st}$ of which is positive; and

$$G_{(15)a}{}^b = \tfrac{1}{30} (- \delta^a_b R^{15} + 420\, \delta^a_b R^{13} R_c{}^d R_d{}^c - 105\, \delta^a_b R^{13} R_{cd}{}^{ef} R_{ef}{}^{cd} - 7280\, \delta^a_b R^{12} R_c{}^e R_d{}^c R_e{}^d + 10{,}920\, \delta^a_b R^{12} R_e{}^c R_f{}^d R_{cd}{}^{ef} + 10{,}920\, \delta^a_b R^{12} R_d{}^c R_{ce}{}^{fg} R_{fg}{}^{de} + 910\, \delta^a_b R^{12} R_{cd}{}^{gh} R_{ef}{}^{cd} R_{gh}{}^{ef} - 3640\, \delta^a_b R^{12} R_{ch}{}^{ef} R_{df}{}^{gh} R_{eg}{}^{cd} - 65{,}520\, \delta^a_b R^{11} R_c{}^d R_d{}^c R_e{}^f R_f{}^e + 131{,}040\, \delta^a_b R^{11} R_c{}^f R_d{}^c R_e{}^d R_f{}^e \ldots), \qquad (73)$$

there being 24,085,338,388 additional terms, the $1^{st}$ of which is negative.

In general, the initial terms (up to 10) of expressions for $G_{(p)a}{}^b$ are given by

$$G_{(p)a}{}^b = \tfrac{1}{2p} ((-1)^p\, \delta^b_a R^p + 4\, (-1)^{p+1} \tbinom{p}{p-2}\, \delta^b_a R^{p-2} R_c{}^d R_d{}^c + (-1)^p \tbinom{p}{p-2}\, \delta^b_a R^{p-2} R_{cd}{}^{ef} R_{ef}{}^{cd} + 16\, (-1)^p \tbinom{p}{p-3}\, \delta^b_a R^{p-3} R_c{}^e R_d{}^c R_e{}^d + 24\, (-1)^{p+1} \tbinom{p}{p-3}\, \delta^b_a R^{p-3} R_e{}^c R_f{}^d R_{cd}{}^{ef} + 24\, (-1)^{p+1} \tbinom{p}{p-3}\, \delta^b_a R^{p-3} R_d{}^c R_{ce}{}^{fg} R_{fg}{}^{de} + 2\, (-1)^{p+1} \tbinom{p}{p-3}\, \delta^b_a R^{p-3} R_{cd}{}^{gh} R_{ef}{}^{cd} R_{gh}{}^{ef} + 8\, (-1)^p \tbinom{p}{p-3}\, \delta^b_a R^{p-3} R_{ch}{}^{ef} R_{df}{}^{gh} R_{eg}{}^{cd} + 48\, (-1)^p \tbinom{p}{p-4}\, \delta^b_a R^{p-4} R_c{}^d R_d{}^c R_e{}^f R_f{}^e + 96\, (-1)^{p+1} \tbinom{p}{p-4}\, \delta^b_a R^{p-4} R_c{}^f R_d{}^c R_e{}^d R_f{}^e \ldots), \qquad (74)$$

there being $b_p - 10$ additional terms, the sign of the $1^{st}$ of which equals $(-1)^p$.